**Electrical properties of volcanic ash samples from Eyjafjallajökull and Grímsvötn**

I M Piper[1,2], K L Aplin[3] and K A Nicoll[4]

1. Met Office, FitzRoy Road, Exeter, Devon, EX1 3PD
2. Department of Electrical and Electronic Engineering, University of Bristol, Woodland Road, Bristol BS8 1UB
3. Department of Physics, University of Oxford, Denys Wilkinson Building, Keble Road, Oxford OX1 3RH
4. Department of Meteorology, University of Reading, P.O. Box 243, Earley Gate, Reading RG6 6BB

**Summary**

Volcanic ash is known to charge electrically, producing some of the most spectacular displays of lightning in nature. Here we investigate the electrical characteristics of ash from two different Icelandic volcanoes - Eyjafjallajökull in 2010 and Grímsvötn in 2011. Laboratory tests investigated the charge transferred to a conducting plate due to fall of volcanic ash through an insulating cylinder. Ash from the Eyjafjallajökull eruption was found to charge slightly positively, whilst Grímsvötn ash was substantially negatively charged. Measurement of the volumetric ratio of particle diameters showed the Eyjafjallajökull ash to have a bimodal distribution, and the Grímsvötn ash a monomodal distribution. Previous experiments with single-material particle systems show that smaller particles charge negatively and larger ones positively [e.g. Lacks and Levandovsky 2007]. Since charge is carried by individual particles, the charging is likely to be dominated by the number size distribution, therefore the large negative charge of the Grímsvötn ash is likely to be related to a large number of small particles within the number size distribution of the ash.

**Introduction**

Volcanic eruptions in Iceland in 2010 and 2011 were associated with significant lightning activity in the ash plume, as measured by the UK Met Office's lightning detection network, ATDnet [Bennett et al. 2010]. A number of mechanisms have been put forward to explain the electrification of volcanic plumes, including triboelectric or fractoemission processes at the vent, the 'dirty thunderstorm' mechanism [e.g. Arason et al. 2011], internal radioactivity of the plume [Mather and Harrison 2006, James et al. 2008] in addition to triboelectric charging within the plume. Sustained electrical charging of the Eyjafjallajökull plume was observed 1200km from the volcano [Harrison et al. 2010] which indicates that some charging of the plume is independent of the eruption process. The lightning activity associated with the eruption from Grímsvötn in 2011 was up to 100 times as intense as that associated with the Eyjafjallajökull eruption in 2010. These observations have motivated a series of experiments in which we investigate the triboelectric charging of samples of volcanic ash from the 2010 Eyjafjallajökull and 2011 Grímsvötn eruption.

**Experiment**

Volcanic ash samples were provided by the Iceland Meteorological Office. Ash from the 2010 Eyjafjallajökull eruption was collected at Sólheimaheiði, 22 km from the crater, and ash from the 2011 Grímsvötn eruption was collected 70 km from the crater.

A series of experiments were carried out whereby samples of volcanic ash were released to fall vertically through a cylinder onto a screened metal plate, located close to the bottom of the cylinder. The charge associated with the ash fall was measured by connecting the metal plate to an electrometer which recorded the voltage on the plate. The apparatus consisted of a 1m long cylindrical Perspex tube, 0.25m in diameter, the top of which was sealed with an inverted funnel connected to a loading trap and shutter (made of cardboard to minimise charge generation within the ash). The voltage on the bottom plate was measured with a Keithley 6512 electrometer and logged to a PC via an IEEE-488 interface. Approximately 50g of ash (baked to remove adsorbed water) was loaded into the trap and the shutter released to enable ash to fall under gravity to the bottom of the tube. The tube was mounted on a support frame so that after each ash drop it was rotated to reload the ash [Krauss et al. 2003, Aplin et al. 2011].

To provide vertical resolution in the measurement of ash charge, a displacement current sensor [Nicoll and Harrison 2009] was mounted on the inside of the Perspex cylinder, halfway between the conducting plate and the shutter of the loading plate. This sensor consists of a spherical electrode connected to an electrometer circuit, which measures the voltage on the electrode. Changes in electrode voltage result from charge transfer from either induction or impaction of ash particles.

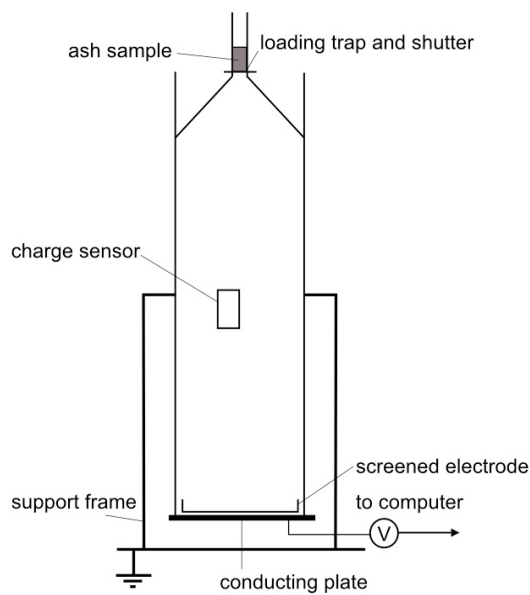

Figure 1: Volcanic ash experiment showing ash loaded before a drop. The support frame permits the tube to be rotated to recover the ash for a subsequent drop.

**Results**

Eleven ash drops were obtained with the Eyjafjallajökull ash, and six for the Grímsvötn ash, with typical results shown in Figure 2. The difference in polarity and magnitude of the voltage generated on the bottom plate between the two ash samples is clear.

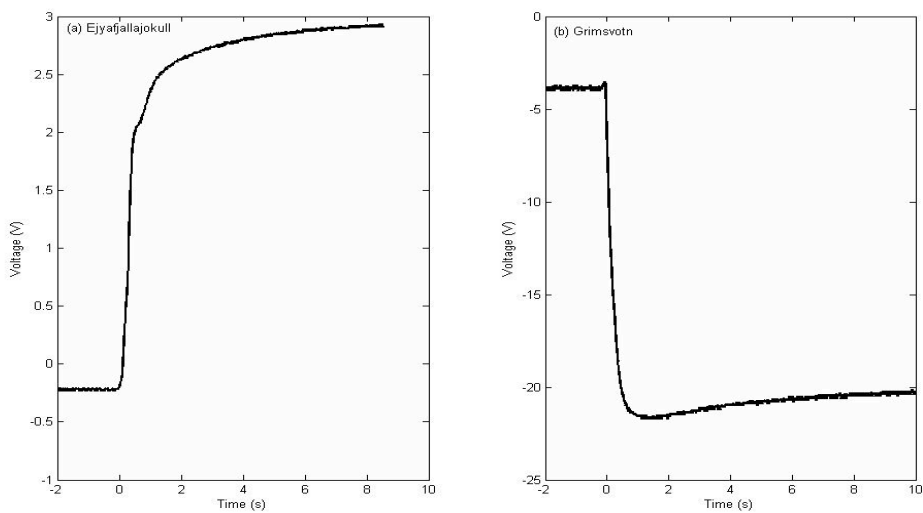

Figure 2: Typical results from ash drops. (a) shows a potential change of + 3.15V for the ash from Eyjafjallajökull, (b) the Grímsvötn ash produces a potential change of -18V.

The capacitance of the system has been estimated to be 100pF using a technique described in Aplin and Harrison (2001). The total charge transferred per unit mass can therefore be estimated to be between +4 to 5.2pC/g for the Eyjafjallajökull ash and -36 to -60pC/g for the Grímsvötn ash. The volumetric distribution of particle diameters in the two samples was measured with a Malvern Mastersizer, and is shown in Figure 3. Although the bulk statistics of each sample were similar, with a median size of 106 and 103 μm for Eyjafjallajökull and Grímsvötn respectively, the Grímsvötn ash shows a monomodal distribution of particle sizes, and Eyjafjallajökull shows a bimodal particle size distribution with peaks at 45μm and 350μm. Grímsvötn had proportionately more small particles by volume, whereas Eyjafjallajökull had more large particles.

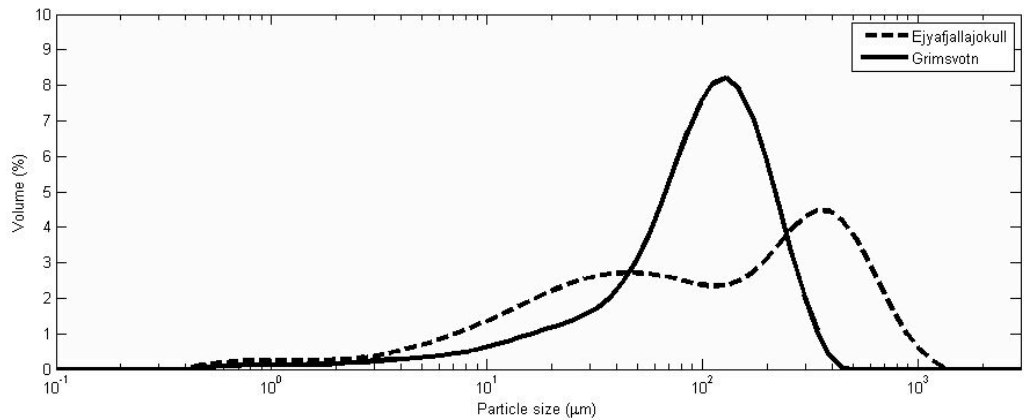

Figure 3: The volumetric particle size distribution for the sample of ash from Eyjafjallajökull is bimodal, whereas the distribution for ash from Grímsvötn is monomodal.

As the median particle sizes were so similar, the enhanced, negative, charge transfer per unit mass for the Grimsvotn sample compared to Eyjafjallajökull is thought to be related to the different sample size distributions.

**Conclusions**

Experiments dropping ash onto a collector plate have indicated that ash from the Grímsvötn eruption is more triboelectrically negative than ash from the Eyjafjallajökull eruption. It is known empirically that in single-material particle systems, the smaller particles charge negatively and the larger ones positively. Lacks and Levandovksy (2007) have explained this phenomenon in terms of transfer of electrons in trapped high-energy states. Smaller particles become depleted of these electrons more quickly than larger particles, which leads to net charge transfer of electrons to the smaller particles from the larger ones. The two peaks in the Eyjafjallajökull sample were relatively similar in size, therefore substantial net charging is not expected. Since charge is carried by individual particles, the charging is likely to be dominated by the number size distribution, rather than the volumetric distribution shown in Figure 3. The number size distribution would show enhanced numbers of small particles for Grimsvotn ash, which would be expected to charge negatively as a consequence.

**Acknowledgements**

We thank the Icelandic Meteorological Office for providing us with the samples of volcanic ash. The University of Oxford Geography Department assisted with the particle size measurements.

**References**

Aplin K L and Harrison R G  2001, A self-calibrating programmable mobility spectrometer for atmospheric ion measurements, *Rev. Sci. Instrum.*, **72**, 8, 3467-3469
Aplin K L, Davis C J, Bradford W J and Herpoldt K L 2011 Measuring Martian Lightning, *J. Phys. Conf. Ser.* **301** 012007
Arason P, Bennett A J and Burgin L E 2011 Charge mechanism of volcanic lightning revealed during the 2010 eruption of Eyjafjallajökull, *J. Geophys. Res.* **116** B00C03
Bennett A J,  Odams P, Edwards D and Arason P 2010 Monitoring of lightning from the April-May 2010 Eyjafjallajökull volcanic eruption using a very low frequency lightning location network, *Environ. Res. Lett.* **5** 044013
Harrison R G, Nicoll K A, Ulanowksi Z and Mather T A 2010 Self-charging of the Eyjafjallajökull volcanic ash plume *Environ. Res. Lett.* **5** 024004
Krauss C E, Horanyi M and Robertson S 2003 Experimental evidence for electrostatic discharging of dust near the surface of Mars *New J. Phys.* **5** 70.1-70.9
Lacks D J and Levandovsky A, 2007, Effect of particle size distribution on the polarity of triboelectric charging in granular insulator systems, *J. Electrostatics.*, 65, 107-112
James M R, Wilson L, Lane S J , Gilbert J S, Mather T A, Harrison R G and Martin R S 2008 Electrical charging of volcanic plumes, *Space Science Reviews* **137** 399-418
Mather T A and Harrison R G 2006 Electrification of volcanic plumes, *Surveys in Geophysics* **27** 4 387-432
Nicoll K A and Harrison R G 2009 A lightweight balloon-carried cloud charge sensor, *Rev. Sci. Inst.* **80** 014501